\newcommand{\AUTHORS}{\
{\centering
{\large Ola Bostr{\"o}m,\footnote{Email address:
tfeob@fy.chalmers.se.}}\raisebox{1ex}{,a}
{\large Mark Miller\footnote{Email address:
mamiller@rodan.syr.edu.}}\raisebox{1ex}{,b}  and
{\large Lee Smolin
\vspace*{5mm}
\footnote{Email address:
smolin@phys.psu.edu.}}\raisebox{1ex}{,c}
\\
{\sl \raisebox{1ex}{a}Institute of Theoretical Physics,
   Chalmers University of Technology,\\ S-412 96 G\"oteborg,
Sweden \\ }
{\sl \raisebox{1ex}{b} Department of Physics, Syracuse University,
 Syracuse, U.S.A., 13244 \\ }
\vspace*{2mm}
{\sl \raisebox{1ex}{c}
Center for Gravitational Physics and Geometry,
Pennsylvania State \\University,
University Park, PA 16802-6300}}}

\documentstyle{article}
\newcommand{\blankline}{\vskip .3cm}
\newcommand{\f}{\begin{equation}}
\newcommand{\ff}{\end{equation}}
\begin{document}
\begin{flushright}
        G\"{o}teborg ITP 94-5 \\
        SU-GP-93-4-1 \\
        CGPG-94/3-3 \\
\end{flushright}
\vfill
\centerline{\LARGE A new discretization of classical and quantum}
\blankline
\centerline{\LARGE general relativity}
\rm
\vskip1cm
\begin{center}
\AUTHORS
\end{center}
 \vfill
\centerline{May 9, 1994}
\vfill
\centerline{Abstract}
\noindent
We propose a new discrete approximation to the Einstein equations, based on
the Capovilla-Dell-Jacobson form of the action for the Ashtekar variables.
This formulation is analogous to the Regge calculus in that the curvature has
support on sets of measure zero.  Both a Lagrangian and Hamiltonian
formulation are proposed and we report partial results about the constraint
algebra of the Hamiltonian formulation.  We find that the discrete versions
of the  diffeomorphism constraints do not commute with each other or with
the Hamiltonian constraint.
\blankline

\vfill\eject

\section{Introduction}

Einstein's equations are beautiful because they capture the
phenomena of
gravitation in
a simple geometrical statement: the vanishing of the Ricci curvature
in
empty regions of
spacetime.  However, when we use them to solve a practical problem
in
gravitation we discover
they have another side---they are complicated nonlinear partial
differential equations.
In special cases, most typically when symmetries have been
imposed, one
sometimes can make
use of the geometry to help discover the solution to a physical
problem.
But, when one is studying the generic problem of constructing
solutions to,
or evolving, the
Einstein equations, little of the geometric beauty comes out in the
techniques we use to try
to solve the theory.

This is especially the case for numerical approximation methods.
Such
methods are crucial
for making progress with important astrophysical questions such as
gravitational wave
production by realistic sources.  By the time all of the elements
necessary
for making the numerical problem well defined are in place,
including
gauge fixing and
finite differencing schemes, very little of the geometrical beauty of
the
equations remains.

For over thirty years there has been available an alternative to the
finite differencing
approaches to the Einstein equations, which is the Regge
calculus\cite{Regge}. In this
approach a
large and generic set of solutions is constructed by limiting
solutions to manifolds
in which the curvature is restricted to have support on sets of
measure
zero.    Typically,
the manifold is broken up into simplices, and the curvature is
restricted to lie on the boundaries at which simplices are joined.

The idea of Regge calculus is that such simplicial manifolds could
approximate a given smooth manifold arbitrarily
well.  Unfortunately, at least up until this
time, Regge
calculus has not been developed into a powerful tool for use in
realistic
calculations.  Or, at least, one should say that this is the case in the
classical theory,
as recently a version of Regge calculus has been shown to yield a
very
effective method
for calculating the path integrals in quantum gravity in
two\cite{JN1986,dynamical,dynamical-2,Forster1987a,Forster1987b},
three\cite{Ponzano-Regge}
and even four
dimensions\cite{migdal-4,Ambjorn,berndenzo}\footnote{For other
work on four dimensional quantum gravity using Regge calculus,
see \cite{RW1981,RW1984,Hamber1990,Hamber1991,Hartle1986}.}.

In this paper we introduce a new discretization of Einstein equations
that
results from applying the exact Einstein's equations to
a
special
restricted class of geometries.   This new formulation has two basic
features.  First,
it is based on the Ashtekar formulation of general relativity
\cite{Abhay,Abhay-perspectives,review-Abhay},
in which the
dynamical
variables are frame fields and self-dual connections.  Second, the
restricted class of
geometries we study are those in which all the fields are
distributional.

These are connected in the following way.  As the Ashtekar
formalism is
polynomial at both
the Lagrangian and Hamiltonian level,
the field equations allow a wider class of solutions than the
conventional
form of the
Einstein equations.  These include solutions in which the determinant
of
the metric
vanishes, for which the usual relations for the Christoffel symbols in
terms
of the
metric components would not be defined.  This is possible because
one of the basic fields of the formalism is a three dimensional
frame field, which is chosen to have density weight one.
(Actually, such solutions are
allowed for all
first order formulations of the Einstein's equation, as has been
emphasized
recently
by Horowitz \cite{Horowitz}.)

Among such degenerate solutions are those for which the densitized
frame field
is actually
distributional.  Because of the Yang-Mills like gauge invariance of the
theory, it turns
out that there are many solutions in which the frame fields have
support
on one dimensional
curves in the Hamiltonian formalisms. This is because the Gauss's law
constraint is solved
by frame fields which are covariantly divergence free.
These configurations may be taken to be  of the form\footnote{We
use throughout index conventions of reference \cite{review-Abhay}.
Early latin
indices ($a,b,c,...$) are three dimensional spatial indices, middle latin
indices ($i,j,k,...$) are three dimensional frame indices, tilde's indicate
density weight one and capital middle latin indices $I,J<...$ refer to
individual loops in a set of loops.  Greek letters are generally used to
denote loops.}
\f
\tilde{E}^{ai}_\alpha (x) =a^2 \int ds \delta^3(x,\alpha (s))
\dot{\alpha}^a (s)  e^i_\alpha
\ff
Here $\alpha $ is a closed or open curve in the spatial three
manifold,
which we will denote
$\Sigma$ and $e^i$ is an element of the Lie algebra of $SU(2)$ which
is
associated with
that curve.  $a$ is a constant with dimensions of length, which is
necessary if, as is natural,
the Lie algebra element $e^i_\alpha$ is dimensionless, so that the
frame
field can also be
dimensionless.

In the Ashtekar formalism it is always important to keep track of the
density weights.  Thus,
note that the right hand side of (1) is naturally a vector density, as is
required.

We can consider
distributional geometries of the form
of (1) for
complicated graphs or lattices.  For example, let $\Gamma$ be some
graph
in $\Sigma$,
with edges $\gamma_I$ where $I=1,...N$.  Then we  consider
distributional configurations
of the form
\f
\tilde{E}^{ai}_\Gamma (x) =a^2 \sum_I \int ds \delta^3(x,\gamma_I
(s))
\dot{\gamma}^a_I (s)  e^i_I
\ff
Associated with the graph $\Gamma$ is then a subspace of the
configuration space of
the theory which is then given by the $N$ lie algebra elements
$e^i_I$.

Now, such a distributional frame field may not seem very physical.
Indeed,
that was the
first impression when expressions of this form first arose in the
quantum
theory in
\cite{tedlee,carlolee}.  However,
we have recently understood\cite{weaves,review-carlo,review-lee}
that such distributional frame fields can actually approximate
arbitrarily well any smooth metric, $q_{ab}$, as long as the scale of
the
curvature of the metric is
large compared to the spacing between the links of the graph, where
both
distances are
measured in terms of $q_{ab}$.

Let us sketch here briefly the main ideas behind this, for more
details the
reader is referred
to \cite{weaves,review-carlo,review-lee}.  Certainly, many
observables on
the configuration
space of general relativity cannot be defined on configurations of the
form
of (2).  Included
in these is the metric at a point.  Since the metric is defined through
the
expression
\f
\tilde{\tilde{q}}^{ab}(x)=\tilde{E}^{ai}(x)\tilde{E}^b_i(x)
\ff
it is not, at least naively, well defined for configurations of the form
of (2)
as the product
of the distributions is not defined.  Moreover, it can be shown that
the
situation cannot
be improved by any regularization or renormalization procedure of
the type
that is usually
used in quantum field theory.  The problem is that any such
procedure
either introduces
spurious dependence on the regularization procedure or changes the
density
character of the
observable\cite{review-lee,review-carlo}.

However, neither in classical nor in quantum physics do we ever
observe the
metric at a
point of space or spacetime.  Leaving aside the difficult issue of
diffeomorphism
invariant observables in general relativity, it is obvious that what is
really observed
in any real experiment in
general relativity are averages of the metric, smeared over some
regions
of space or spacetime.

The question is then whether there are observables which measure
the
metric smeared over
regions of space or spacetime which can be defined on distributional
configurations of
the form of (2).  It turns out that there are a number of such
observables.
Among them
are the following three \cite{weaves,review-lee}:

1)  The area of any given two dimensional surface,
$\cal S$, in $\Sigma$.  It is denoted $\hat{\cal A}[{\cal S}]$.

2)  The volume of any three dimensional region
$  \cal R$ in $\Sigma$.  It is denoted $\hat{\cal V}[{\cal R}]$.

3)  An observable that  measures the integrated norm of any one
form
$\omega$
on $\Sigma$.  Written in terms of the
classical three metric $q_{ab}$ it is
\f
Q[\omega ] = \int_\Sigma \sqrt{det(q)q^{ab}\omega_a \omega_b }  .
\ff
Note that the square root is a density and is thus integrable.

The key point is  that in spite of the distributional character of the
frame fields
(2) the areas and volumes associated with them are finite when they
are
nonvanishing.  For
example, the area that  a surface  $\cal S$ acquires from the
distributional configuration
$\tilde{E}^{ai}_\Gamma$ is nonvanishing as long as the surface
intersects
the graph at
least once.  In that case the area is
\f
{\cal A}[{\cal S}] = a^2 \sum_{I}{\cal I}^+[{\cal S},\gamma_I] |e_I|
\ff
where $|e_I|$ is the norm of the $SU(2)$ Lie algebra and
${\cal I}^+[{\cal S},\gamma_I]$ is the unoriented intersection number,
that
simply counts positively the number of intersections of the surface
and
the curve.
The area observable will be described in detail in the appendix.

Thus, if we are measuring, not the metric at a point, but the areas of
surfaces, it is
possible to arrange the graph and choose the $e_I^i$ such that the
areas
the surfaces
have in the distributional configuration (2) approximate arbitrarily
well the
areas they
have given a
 smooth metric $q_{ab}$.  For example, one way to do this
is the
following\cite{weaves}.  Fix $|e_I|=1$.  Then space the lines in the
graph so that, on the
average
one line crosses every surface once per $a^2$ units of area of the
surface,
as it is
measured by $q_{ab}$.  This can be done consistently for any set of
surfaces, as
long as their radii of curvature are small compared to $a$, where
again,
the radii of
curvatures are measured with respect to $q_{ab}$.  Then, consider
any
such surface
whose area from $q_{ab}$ is large compared to $a^2$.  Its area
according
to the
distributional geometry (2), from (5), will then be equal to the value
from the
smooth
metric, up to errors of order $a^2$ divided by that area.

Similar statements can be made concerning the second and third
observables.  We may conclude
that as long as we only measure such smeared observables, any
smooth
three metric may be
approximated arbitrarily well by distributional metrics of the form
of (2).

Having established this correspondence between distributional and
smooth
three metrics, we may go on to ask four additional questions.
First, can we define a
corresponding set of
distributional connections such that the constraints of general
relativity
can be extended to the case of such distributional connections and
frame fields?   Second, can we
find solutions to the constraints among these distributional initial
data, yielding an extension of the physical phase space of general
relativity to such distributional solutions.   Third, can we define
Poisson
brackets on the space of distributional frame and connection fields
that,
together with the extention of the constraints to these fields, defines
a constrained Hamiltonian system?  Fourth, can we find
corresponding
 distributional solutions to the full set of Einstein's equations, either
by evolving the
distributional solutions to the constraints in time or
by directly solving
the Einstein equations.

The
purpose of this paper is to show that the answer to the first three
questions is yes.   We find that the equations of general relativity
can be consistently reduced to a set of equations that govern the
evolution of distributional frame and connection fields.  The starting
point for this formulation of dynamics is to begin with a
form of
the action first written down by
Capovilla, Dell and Jacobson\cite{CDJ}.  This allows
us to make a consistent
truncation of the constraint equations onto a finite, but arbitrarily
large,
dimensional
phase space, that can approximate arbitrarily well smooth solutions
to the
constraints of general relativity.  As in full general relativity,
the Hamiltonian is a linear combination of constraints, and those
constraints are in correspondence with the constraints of the
continuum
theory so that there are discrete analogues of the Hamiltonian,
diffeomorphism and $SU(2)$ gauge constraints.

The key to these results is to invent connection fields that are
associated with the faces of the graph $\Gamma$.  If we label the
faces by
an
index $\alpha$, the connection will also be specified by associating a
Lie
algebra
$a_\alpha$ to each face.  Associated to every graph $\Gamma$ we
will
then have a
phase space ${\cal P}_\Gamma$, in which $SU(2)$ Lie algebra
elements
corresponding to frame fields of the form of
(2) are associated to each link and the conjugate fields are Lie
algebra
elements associated to each face.

Given a formulation of a discrete approximation to general
relativity as a constrained Hamiltonian system, we would
like to use its evolution equations to construct solutions
that approximate smooth solutions to general relativity.
In order to do this, the key question which must be answered
is that of the  algebra of the constraints
of the discrete theory.  One's first expectation is that the constraint
algebra is most likely second class, as a result of the fact that the
diffeomorphism invariance of the continuum theory appears to be broken
by the reduction to the discrete theory.  This is known to occur in
several attempts to construct  Hamiltonian formulations of the
Regge calculus
\cite{johnian,paullee,Waelbroeck1989,Waelbroeck1990}.
We have studied the constraint algebra in the
present model and we report here partial results concerning its form.
While we have not yet been able to calculate the full algebra, we
have
been able to compute the algebra in a certain limit, in which the
$SU(2)$
internal gauge symmetry is reduced to an abelian algebra which is
$U(1)^3$.  In the continuum theory this limit has been studied, and
corresponds to the limit in which Newton's constant $G$ is taken
to zero in the Ashtekar formalism\cite{Gtozero}.  It corresponds to a
chirally
asymmetric theory which includes the full self-dual sector of the
theory but only the
linearization of the antiself-dual sector.
What we find is that in this limit the algebra of the constraints
analogous
to the gauge and spatial diffeomorphism constraints is not first
class
in this discrete approximation.

Finally, it may also be interesting to construct discrete solutions to
general relativity which approximate smooth solutions by
restricting the four dimensional equations of motions of the theory
to distributional fields so that time, as well as space, becomes
discrete.    We show in
section 4 below that there is such a formulation, which is based as
well
on the Capovilla-Dell-Jacobson form of the action.

This paper is organized as follows.  In the next section we derive
the discrete Hamiltonian formulation by reducing the
Capovilla-Dell-Jacobson
 form of the
action for general relativity to a form appropriate for fields that are
distributional on three surfaces, but continuous in time.  In section 3
we
study the algebra of constraints of the resulting Hamiltonian
formulation,
while section 4 is devoted to the construction of a four dimensional
lagrangian in which the fields are distributional on the spacetime
manifold.
Our conclusions and suggestions for further work are in
section 5.  The appendix contains technical details
about the extension of the area observable to distributional
fields.

\section{The Hamiltonian formulation}

The best way to insure that a constrained Hamiltonian formulation is
consistent
is to derive it from an action principle.  In this paper we will thus
describe two closely related formulations of an action principle for
distributional fields in general relativity\footnote{A preliminary
version of the
formulation of this section was presented in \cite{review-lee}.
 We may note that
several of the equations of that presentation have been
corrected here, including
the relationship between the connection and curvatures.}.
In the first, spacetime
is discretized,
so that the full four dimensional Einstein equations are
replaced by a finite
set of difference equations.  This is the subject
of section 4 below.  In the second, space is discretized, but time
is kept continuous.  This is necessary if we are to derive a discrete
approximation to the Hamiltonian theory, and is the subject of the
present
section.

Both types of Lagrangians are constructed by using a form of the
Lagrangian for general
relativity first written down by Capovilla, Dell and Jacobson.  It
will be helpful to first explain this form, as it is the starting point
for all the developments of this paper.
On spacetime, which will be denoted ${\cal M}$, we consider two
independent fields.
These are an  $SU(2)$ connection one form, denoted $A^i$ and a
three by
three matrix
of scalar fields, denoted $\phi^i_j$.  We may then consider the action
\f
S(A,\phi ) = \int F^i \wedge F^j [\phi^{-1}]_{ij}
\ff
In this form of the variational principle, $\phi^i_{\ j}$ is to be
varied respecting two constraints.  These are a symmetry condition,
\f
\phi_{ij}=\phi_{ji}
\ff
and a trace condition
\f
\sum_i \phi_{ii} = -6 \Lambda
\ff
where $\Lambda$ is the cosmological constant (which may be zero).

Let us assume that $\cal M$ has the topology of $\Sigma \times R$,
for
some three
manifold $\Sigma$, and consider a $3+1 $ decomposition of $\cal M$.
We
may then  split
the spacetime coordinates, $x^\alpha$ into a time coordinate $t$ and
spatial coordinates
$x^a$ on $\Sigma$.  It is then easy to show that if one defines the
corresponding Hamiltonian
theory for evolution in $t$ the Ashtekar form of the Hamiltonian
theory is
found.

Let us sketch this, as we will shortly be following the same
procedure in
the distributional
case.  We may define the frame fields $\tilde{E}^{ai}$ as the
conjugate
momenta to $A_a^i$, according to the usual
\f
\tilde{E}^{ai} = { \delta S   \over \delta \dot{A_a^i}} =
2[\phi^{-1}]^i_{\ j}
\tilde{B}^{aj}
\ff
where $\tilde{B}^{ai} \equiv {1 \over 2} \epsilon^{abc}F_{bc}^i$.
It is then one line to show that the action may be written as
\f
S= \int dt \int_\Sigma \left [ \tilde{E}^{ai} \dot{A_a^i} - A_0^i
{\cal G}^i
\right ]
\ff
where ${\cal G}^i$ is the Gauss's law constraint defined by
\f
{\cal G}^i(x) = {\cal D}_a \tilde{E}^{ai}(x) = 0
\ff
where ${\cal D}_a$ is the $SU(2)$ gauge covariant derivative.
The
Poisson brackets
may be read off from (10), they are,
\f
\{A_{bj}(x), \tilde{E}^{ai}(y) \} = \delta^3 (x,y) \delta^a_b \delta^i_j
\ff
However, because of the conditions (7) and (8) we put on
$\phi_{ij}$, not all
the
$\tilde{E}^{ai}(x)$ defined by (9) are independent.  Instead there are
primary
constraints.
It is not hard to show that there are exactly the momentum and
Hamiltonian constraints
of the Ashtekar formalism:
\f
{\cal C}_a = F_{ab}^i \tilde{E}^{bi} = 0
\ff
\f
{\cal C} = \epsilon_{ijk} F_{ab}^i \tilde{E}^{aj}\tilde{E}^{bk} +
\Lambda \epsilon_{ijk} \epsilon_{abc} \tilde{E}^{ai}
\tilde{E}^{bj}\tilde{E}^{ck} = 0
\ff
where $F_{ab}^i$ is the Yang-Mills curvature associated with $A_a^i$.

It is intriguing that in this formulation the dynamical constraints of
general relativity
are recovered as {\it primary constraints}.   This will turn out to be a
great help when
constructing the discretization.

Now, we are going to construct a discretization by restricting this
form of
the action to
distributional fields.  The key question to be answered is what kind
of
distributional
connections and curvatures will be employed to do this.  In order to
answer
this
question, let us first note that by (9)
the support of $\tilde{E}^{ai}$ is likely to coincide with the support of
$\tilde{B}^{ai}= {1 \over 2} \epsilon^{abc}F_{bc}^i$.   Thus,
returning for the moment to the three dimensional formalism, we
will
expect to have
curvatures of the form,
\f
\tilde{B}^{ai}_\alpha  (x) = {1 \over G} \sum_I \int ds
\delta^3(x,\gamma_I
(s))
\dot{\gamma}^a_I (s)  b^i_I   .
\ff
Here, the $  G$ is put in for dimensions. We would
like the free factors $b^i_{\vec{n} \hat{a}}$ to be dimensionless.
In Ashtekar's
formalism it is $  GF_{ab}^i$ that
has the dimensions of curvature, which is inverse length squared.

Thus, in the Hamiltonian theory the spacetime self-dual curvature
will be
represented
by one $SU(2)$ Lie algebra element, $b_I^i$, associated to each line of
the
graph.

\begin{figure}[t]
\begin{picture}(400,90)(0,-10)
\put(25,30){\vector(1,0){25}}
\put(25,30){\vector(3,2){15}}
\put(25,30){\vector(0,1){25}}
\put(55,30){$\hat{x}$}
\put(30,55){$\hat{z}$}
\put(45,45){$\hat{y}$}
\put(90,30){$e^i_{\vec{n}\hat{x}}$ =}
\put(205,30){$a^i_{\vec{n}\hat{x}}$ =}
\put(185,35){i}
\put(285,65){i}
\put(250,0){\circle*{5}}
\put(130,30){\circle*{5}}
\put(250,-10){$\vec{n}$}
\put(130,20){$\vec{n}$}
\put(130,30){\vector(1,0){50}}
\put(250,0){\line(0,1){50}}
\put(280,20){\line(0,1){50}}
\put(250,0){\line(3,2){30}}
\put(250,50){\line(3,2){30}}
\end{picture}
\begin{center}\parbox{12cm}{
\caption{The lattice variables $e^i_{\vec{n}\hat{x}}$ and
$a^i_{\vec{n}\hat{x}}$, where
 $\hat{x}$ means
the direction of $e^i_{\vec{n} \hat{x}}$ respectively the normal of the face
${\cal S}_{\vec{n}\hat{z}\hat{y}}$.}
\label{aab}
}\end{center}
\end{figure}
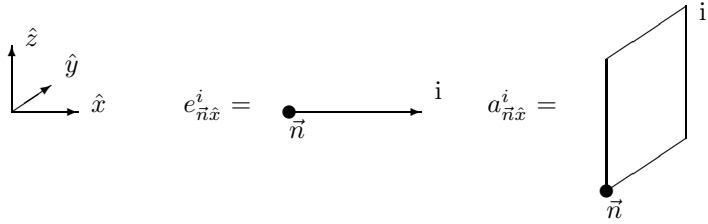

The question we must now ask is what form of a distributional
connection gives such
a curvature?
It turns out that the right answer is that the connection has support
on the
faces of the
graph.
In order to make the combinatorics simple and explicit, for the
rest of this paper
we will restrict attention to cubic lattices.
Thus, given an arbitrary coordinate chart
on $\Sigma$ let us define a standard cubic lattice with coordinate
lattice spacing $a$.  The vertices will be labeled by three integers
$\vec{n}$ and the links by the pair $(\vec{n}, \hat{a})$.
Thus, $\gamma_{\vec{n} \hat{a}} (s)$, with $s \in (0,1)$ will
be taken to refer to the link leaving the vertex $\vec{n}$ in the
positive $\hat{a}$ direction.
{}From now on, $\hat{a}$, $\hat{b}$, $\hat{c}$ will be one of three positive
 directions in a
right-handed system, see fig(\ref{aab}).
The vertices themselves are located at
points $x_{\vec{n}} = \gamma_{\vec{n} \hat{a}} (0)$.

\begin{figure}[t]
\Large
\begin{picture}(500,110)(0,-10)
\put(-10,30){$b_{\vec{n} \hat{a}}^i$ =}
\put(40,30){=}
\normalsize
\put(110,10){\circle*{5}}
\put(35,10){\circle*{5}}
\put(35,00){$\vec{n}$}
\put(110,00){$\vec{n}$}
\put(220,10){\circle*{5}}
\put(220,00){$\vec{n}$}
\put(320,10){\circle*{5}}
\put(320,00){$\vec{n}$}
\put(70,45){--}
\put(125,65){--}
\put(140,45){+}
\put(90,30){+}
\put(350,45){+}
\put(300,45){+}
\put(210,40){+}
\put(240,60){+}
\put(170,60){i}
\put(375,60){k}
\put(260,70){j}
\put(165,30){$+2{\varepsilon}^{ijk}$}
\thicklines
\put(35,10){\vector(0,1){50}}
\thinlines
\put(60,10){\line(0,1){50}}
\put(60,10){\line(1,0){20}}
\put(60,60){\line(1,0){50}}
\put(110,10){\framebox(50,50)}
\put(140,80){\line(0,-1){20}}
\put(80,40){\line(3,2){60}}
\put(80,40){\line(0,-1){50}}
\put(80,-10){\line(3,2){30}}
\put(320,10){\framebox(50,50)}
\put(270,10){\framebox(50,50)}
\put(190,40){\line(3,2){60}}
\put(190,-10){\line(3,2){60}}
\put(190,40){\line(0,-1){50}}
\put(250,80){\line(0,-1){50}}
\put(220,60){\line(0,-1){50}}
\end{picture}
\begin{center}\parbox{12cm}{
\caption{The magnetic component $b_{\vec{n} \hat{a}}^i$, in the direction
$\hat{a}$, in terms of the connections
$a^i_{\vec{n}\hat{a}}$.}
\label{b}
}\end{center}
\end{figure}
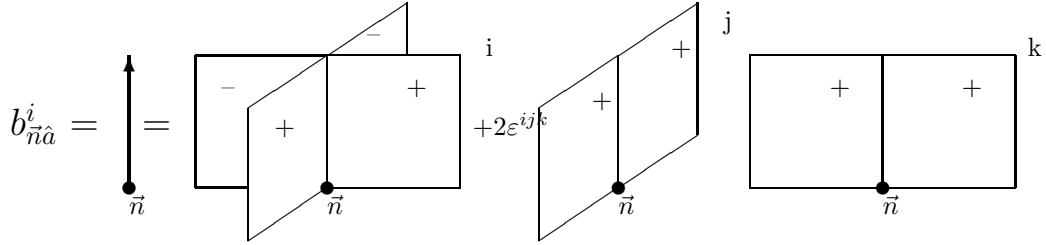

We will then take the frame field and curvature to be of the
forms (2) and (15), with the index $I$ over the links now labeled by
$(\vec{n},
\hat{a})$,

Let us also label the faces by $(\vec{n}, \hat{a}\hat{b})$ with
$\hat{a} >
\hat{b}$.
Then, we may consider distributional connections of the form,
\f
A_a^i (x) =  {1 \over G} \sum_{\vec{n}\hat{a}\hat{b}, \hat{a}>\hat{b}}
\int d^2S^{bc}_{\vec{n}\hat{a}\hat{b}}(\sigma )
\epsilon_{abc}
\delta^3 (x,{\cal S}_{\vec{n}\hat{a}\hat{b}} (\sigma) )
a^i_{\vec{n}\hat{c}}
\ff
where $\hat{c}$ is the positive normal of the face spanned by  $\hat{a}$
and  $\hat{b}$, see fig(\ref{aab}), and
 $\sigma$ stands for an arbitrary pair of  coordinates on the
two dimensional surface.
%%%%%%%%%%%%%%%%%%%% ADDED %%%%%%%%%%%%%%%%%%%%%%%%%%%%%%%%%%%%%%%%%%%%%
%%%%%%%%%%%%%%%%%%%%%%%%%%%%%%%%%%%%%%%%%%%%%%%%%%%%%%%%%%%%%%%%%%%%%%%%%%%%%
We must first compute the associated curvatures and show that they
are
of the form of (15),
so that the $b_{\vec{n}, \hat{a}}^i$ can be expressed in terms of the
$a^i_{\vec{n}\hat{a}}$.  To compute the first, derivative term
of (15)
smear the
distribution with an appropriate one form $f_{ci}$ and compute,
\f
\int d^3x f_{ci} \epsilon^{abc} \partial_a A_b^i =
{1 \over G} \sum_{\vec{n}\hat{a}} \left [
\sum_{\hat{b}\hat{c}} \epsilon^{\hat{a}\hat{b}\hat{c}}
\int ds \dot{\gamma}^a_{\vec{n}\hat{a}} (s)
f_{ai} (\gamma_{\vec{n}\hat{a}} (s) )
\left ( a^i_{\vec{n}\hat{c}} - a^i_{(\vec{n}-
\hat{b})\hat{c}}
\right ) \right ]
\ff
The second, non-abelian term is also found to have support only on
the
links of the lattice.
The result is that the curvature of the connection (16) is of the form
of (15),
with
\f
b^i_{\vec{n}\hat{a}} = \sum_{\hat{b},\hat{c} \neq \hat{a}}
\epsilon^{\hat{a}\hat{b}\hat{c}} \left [
\left ( a^i_{\vec{n}\hat{c}} - a^i_{(\vec{n}-
\hat{b})\hat{c}}
\right )
+ {1 \over 8} \epsilon^{ijk}
\left ( a^j_{\vec{n}\hat{b}} + a^j_{(\vec{n}-
\hat{c})\hat{b}}
\right )
\left ( a^k_{\vec{n}\hat{c}} + a^k_{(\vec{n}-
\hat{b})\hat{c}}
\right ) \right ]
\ff
see also fig(\ref{b}).
%%%%%%%%%%%%%%%%%%%%%%%%%% ADDED %%%%%%%%%%%%%%%%%%%%%%%%%%%%%%%%%%%%%%
%%%%%%%%%%%%%%%%%%%%%%%%%%%%%%%%%%%%%%%%%%%%%%%%%%%%%%%%%%%%%%%%%%%%%%%

In the continuum, the uncontracted Bianchi identity gives a relation
between the curvature $\tilde{B}^{ai}$ and the connection $A^i_a$,
namely
\f
{\cal D}_{[a} ( \epsilon_{bc]d} \tilde{B}^{di} ) = 0 \label{Bianchi}
\ff
It is easy to verify this equation directly by substituting
\f
\tilde{B}^{ai} = \epsilon^{abc} F_{bc}^i =
\epsilon^{abc} ( \partial_b A^i_c - \partial_c A^i_b +
\epsilon^{ijk} A_b^j A_c^k )
\ff
into the left-hand side of (19).
Substituting the distributional fields (15) and (16) into the left-hand
side of (19) yields
\begin{eqnarray}
{\cal D}_{[a} ( \epsilon_{bc]d} \tilde{B}^{di} ) = & {\sum_{\vec{a}}} &
\left [ b^i_{\vec{n}\hat{a}} - b^i_{(\vec{n}-\hat{a})\hat{a}}
\right.
\\ \nonumber
&& \left.
 +
{1 \over 8}\epsilon^{ijk} \left ( a^j_{\vec{n}\hat{a}} +
{\sum_{\hat{b},\hat{c} \neq \hat{a}\;\;\hat{b}\neq \hat{c}}}
[{1 \over 2}a^j_{(\vec{n}-(\hat{b}+\hat{c}))\hat{a}}
+ a^j_{(\vec{n}-\hat{b})\hat{a}}
] \right ) \left (
b^k_{\vec{n}\hat{a}} + b^k_{(\vec{n}-\hat{a})\hat{a}}
\right )  \right ]
\end{eqnarray}
One might expect that substituting (18) into (21) would make the
expression identically zero.  This turns out not to be the case.
The right hand side of (21), expressed purely in terms of
$a^i_{\vec{n}\hat{a}}$ via (18), has terms that are
linear, quadratic, and cubic in $a^i_{\vec{n}\hat{a}}$.
Only the linear and quadratic terms cancel, leaving an expression
cubic in $a^i_{\vec{n}\hat{a}}$ that, in general, is not zero.

To understand this, we remember that the Bianchi identity can be interpreted
geometrically as a manifestation of the "boundary of a boundary = 0"
principle\cite{MTW}.  Taking a cube in the dual lattice with lattice
vertex $\vec{n}$ at the center, we parallel transport a test spinor
around each of its faces in such a way that each link gets traversed
twice, once in each direction.  The resulting change in the test spinor
is, to lowest non-zero order in $b^i_{\vec{n}\hat{a}}$, exactly
given by the linear and quadratic (in $a^i_{\vec{n}\hat{a}}$)
part of the right hand side of (21).  Thus, the Bianchi identity for
our distributional frame fields is
\begin{eqnarray}
&&{\sum_{\vec{a}}}[
b^i_{\vec{n}\hat{a}} - b^i_{(\vec{n}-\hat{a})\hat{a}}
\\&& \nonumber
 +
{1 \over 8}\epsilon^{ijk} \left ( a^j_{\vec{n}\hat{a}} +
{\sum_{\hat{b},\hat{c} \neq \hat{a}\;\;\hat{b}\neq \hat{c}}}
[ {1 \over 2}a^j_{(\vec{n}-(\hat{b}+\hat{c}))\hat{a}}
+ a^j_{(\vec{n}-\hat{b})\hat{a}}
] \right ) \left (
\beta^k_{\vec{n}\hat{a}} + \beta^k_{(\vec{n}-\hat{a})\hat{a}}
\right )]  = 0 \label{Bianchi2}
\end{eqnarray}
where
\f
\beta^i_{\vec{n}\hat{a}} = {\sum_{\hat{b},\hat{c} \neq \hat{a}}}
\epsilon^{\hat{a}\hat{b}\hat{c}}
(a^i_{\vec{n}\hat{c}} - a^i_{(\vec{n}-\hat{b})\hat{c}})
\ff
That is, $\beta^i_{\vec{n}\hat{a}}$ is equal to the linear terms
of $b^i_{\vec{n}\hat{a}}$ in (18).

This completes our specification of the kinematical structure of the
discretization.   The next question to be addressed is how to extend
the
constraint equations to distributional fields of the form of (2) and
(16).  This
is not a completely trivial problem, as the computations involve
products
of distributions.  For this reason, the simplest approach is to use the
Lagrangian approach to define the dynamics.

In the continuum theory, the action can be written
\f
S(A_a,A_0,\phi) = 2 \int dt \int_\Sigma
\left [ \dot{A^i_a} - {\cal D}_a A^i_0 \right ]
\tilde{B}^{aj} \phi^{-1}_{ij}
\ff

We now make the coefficients of the distributional fields,
$e^{i}_{\vec{n}\hat{a}}$ and $a^{i}_{\vec{n}\hat{a}}$, time
dependent and plug these fields into the above action.  When
we do this, we notice that in order to perform the
integrations, some kind of restriction has to be put on
$\phi_{ij}$.  To see what restriction we should choose, notice
that in the Ashtekar formalism there are 18 configuration
variables at each point in $\Sigma$ ; $\tilde{E}^{ai}$ and
$A^i_a$ (we do not count $A^i_0$ as a configuration variable,
since it plays the role of a Lagrange multiplier).
We would therefore expect to have 18 configuration
variables associated with each vertex of the lattice.  This turns
out to be the case if we restrict $\phi_{ij} (x)$ to be constant
on the three links associated with each vertex.
That is, $\phi_{ij}(x)$ has the same value, which we will denote
as $\phi_{ij}(x_{\vec{n}})$, on the three links obtained by starting
at the vertex $\vec{n}$ and moving in the three positive directions.
Note that $\phi_{ij}(x)$ is discontinous at each vertex.  We now have
18 configuration variables associated with each vertex (9 a's and
9 $\phi$'s).

If we make this restriction, keeping $A^i_0 (x)$ unrestricted, then
the discrete version of the action is
\f
S = \int dt {\cal L}[a,A_0,\phi]
\ff
where
\begin{eqnarray}
{\cal L}[a,A_0,\phi ]& = &{2 \over {8 G^2}}
\sum_{\vec{n},\hat{c}}
\dot{a}_{\vec{n}\hat{c}}^i  \times \left \{
b^i_{\vec{n}\hat{c}} \phi^{-1}_{ij}(x_{\vec{n}})
+b^i_{(\vec{n}-\hat{c})\hat{c}} \phi^{-1}_{ij}(x_{\vec{n}-\hat{c}})
\right.
\\  \nonumber
&& \left.
+\sum_{\hat{a},\hat{b}\neq \hat{c}\;\;\hat{a}\neq \hat{b}}[{1 \over 2}
 b^i_{(\vec{n}+
\hat{a}+\hat{b})\hat{c}} \phi^{-1}_{ij}(x_{\vec{n}+
\hat{a}+\hat{b}})+ b^i_{(\vec{n}+
\hat{a})\hat{c}} \phi^{-1}_{ij}(x_{\vec{n}+
\hat{a}})
+
\right.
\\  \nonumber
&& \left.
+ {1 \over 2}
 b^i_{(\vec{n}-\hat{c}+
\hat{a}+\hat{b})\hat{c}} \phi^{-1}_{ij}(x_{\vec{n}-\hat{c}+
\hat{a}+\hat{b}})+ b^i_{(\vec{n}-\hat{c}+
\hat{a})\hat{c}} \phi^{-1}_{ij}(x_{\vec{n}-\hat{c}+
\hat{a}})]
\right. \} +
\\   \nonumber
&&   + {1 \over G} \sum_{\vec{n}} A_0^i (x_{\vec{n}})
{\cal G}^i_{\vec{n}}
\end{eqnarray}
Here, ${\cal G}^i_{\vec{n}}$ is  the discrete version
of the Gauss law constraint; it is given by,
\begin{eqnarray}
{\cal G}^i_{\vec{n}}&=2 \sum_{\hat{a}} & \left  (
b^j_{\vec{n}\hat{a}} \phi^{-1}_{ij} (x_{\vec{n}}) -
b^j_{(\vec{n}-\hat{a})\hat{a}} \phi^{-1}_{ij} (x_{\vec{n}-\hat{a}})
\right )
\\   \nonumber
&&  +  {1 \over 8}\epsilon^{ikl} \left (
b^j_{\vec{n}\hat{a}} \phi^{-1}_{lj} (x_{\vec{n}}) +
b^j_{(\vec{n}-\hat{a})\hat{a}} \phi^{-1}_{lj} (x_{\vec{n} - \hat{a}})
\right )
\\   \nonumber
&& \times \left (
a^k_{\vec{n}\hat{a}} + \sum_{\hat{b},\hat{c}\neq \hat{a}\;\;\hat{b}\neq
\hat{c}}
[ {1 \over 2}a^k_{(\vec{n}-(\hat{b}+\hat{c}))\hat{a}}+
a^k_{(\vec{n}-\hat{b})\hat{a}}
] \right )
\end{eqnarray}

Following the continuum theory, let us define the coefficients
of the frame field to be

\f
e_{\vec{n}\hat{a}}^j \equiv 2\phi^{-1}_{ij} (x_{\vec{n}})
b_{\vec{n}\hat{a}}^i \label{fi}
\ff
The Lagrangian is now
\begin{eqnarray}
{\cal L}[a,A_0,e]&= & {1 \over {8 G^2}}
\sum_{\vec{n},\hat{c}}
\dot{a}_{\vec{n}\hat{c}}^i  \times \left \{
e^i_{\vec{n}\hat{c}}
+e^i_{(\vec{n}-\hat{c})\hat{c}}
\right.
\\  \nonumber
&& \left.
+\sum_{\hat{a},\hat{b}\neq \hat{c}\;\;\hat{a}\neq \hat{b}}[ {1 \over 2}
 e^i_{(\vec{n}+
\hat{a}+\hat{b})\hat{c}} +
e^i_{(\vec{n}+
\hat{a})\hat{c}}
+
\right.
\\  \nonumber
&& \left.
+{1 \over 2}
 e^i_{(\vec{n}-\hat{c}+
\hat{a}+\hat{b})\hat{c}} + e^i_{(\vec{n}-\hat{c}+
\hat{a})\hat{c}}]
\right. \} +
\\   \nonumber
&&   + {1 \over G} \sum_{\vec{n}} A_0^i (x_{\vec{n}})
{\cal G}^i_{\vec{n}}
\end{eqnarray}

Now, the Gauss constraint is
\begin{eqnarray}
{\cal G}^i_{\vec{n}} = \sum_{\hat{a}}
\left (
e^i_{\vec{n}\hat{a}} - e^i_{(\vec{n}-\hat{a})\hat{a}}
\right ) + \\\nonumber
{1 \over 8} \epsilon^{ijk} \left (
e^k_{\vec{n}\hat{a}} + e^k_{(\vec{n}-\hat{a})\hat{a}} \right )
&& \times \left (
a^k_{\vec{n}\hat{a}} + \sum_{\hat{b},\hat{c}\neq \hat{a}\;\;\hat{b}\neq
\hat{c}}
[{1 \over 2}a^k_{(\vec{n}-(\hat{b}+\hat{c}))\hat{a}}+
a^k_{(\vec{n}-\hat{b})\hat{a}} ]
\right )
\end{eqnarray}

Notice that, at each vertex, we have replaced the five independent
variables $\phi_{ij} (x_{\vec{n}})$ with the nine variables
$e^i_{\vec{n}\hat{a}}$.  Due to conditions (7) and (8) on
$\phi_{ij}$, we have 4 primary constraints on $e^i_{\vec{n}\hat{a}}$
at each vertex:
\f
{\cal C}_{\vec{n}\hat{a}}= \epsilon_{\hat{a}\hat{b}\hat{c}}
 e^i_{\vec{n} \hat{b}} b^i_{\vec{n} \hat{c}} \label{ve}
\ff
\f
{\cal C}_{\vec{n}} = \epsilon_{ijk} \epsilon_{\hat{a}\hat{b}\hat{c}}
e^i_{\vec{n} \hat{a}}  e^j_{\vec{n} \hat{b}} b^k_{\vec{n} \hat{c}}  \label{sc}
\ff
These are the lattice vector and scalar constraints,
respectively.
%%%%%%%%%%%%%%%%%%%%%%% ADDED %%%%%%%%%%%%%%%%%%%%%%%%%%%%%%%%%%%%%%%%%

In the continuum, the abelian term of the vector constraint is equal to
the abelian term of the Gauss constraint times the connection. The physical
interpretation of this is that the diffeomorphism constraint vanishes in the
abelian limit when the $G_{Newton}$ constant goes to infinity \cite{Husain}.
This is not the case for the discretized constraints (\ref{ve}).

%%%%%%%%%%%%%%%%%%%%%%%%%%%%%%%%%%%%%%%%%%%%%%%%%%%%%%%%%%%%%%%%%%%%%%%

\section{The constraint algebra}

We would now like to find the Poisson bracket algebra of the
constraints.
Notice that the definition of the momenta conjugate to both
$e^i_{\vec{n}\hat{a}}$ and $a^i_{\vec{n}\hat{a}}$ are actually
primary constraints, since they do not involve time derivatives of
$e^i_{\vec{n}\hat{a}}$ and $a^i_{\vec{n}\hat{a}}$ :

\begin{eqnarray}
{\Pi}_{a^i_{\vec{n}\hat{c}}} &= &
{d{\cal L} \over d{\dot{a}}^i_{\vec{n}\hat{c}}} =
{1 \over 8} \{ e^i_{\vec{n}\hat{c}}
+e^i_{(\vec{n}-\hat{c})\hat{c}}
\\  \nonumber
&& \left.
+\sum_{\hat{a},\hat{b}\neq \hat{c}\;\;\hat{a}\neq \hat{b}}[{1 \over 2}
 e^i_{(\vec{n}+
\hat{a}+\hat{b})\hat{c}} +
e^i_{(\vec{n}+
\hat{a})\hat{c}}
+
\right.
\\  \nonumber
&& \left.
+{1 \over 2}
 e^i_{(\vec{n}-\hat{c}+
\hat{a}+\hat{b})\hat{c}} + e^i_{(\vec{n}-\hat{c}+
\hat{a})\hat{c}}] \} \right.
\end{eqnarray}

\f
{\Pi}_{e^i_{\vec{n}\hat{a}}} =
{d{\cal L} \over d{\dot{e}}^i_{\vec{n}\hat{a}}} = 0
\ff

Let us label these constraints by
\f
{\chi}_{I} = 0
\ff
where I is an index labeling the above constraints.
%%%%%%%%%%%%%%%%%%%% ADDED %%%%%%%%%%%%%%%%%%%%%%%%%%%%%%%%%%%%%%%%%%%
Let us now treat a $L\times M\times N$ cubic
lattice, with opposite points on the boundary identified, so
that the topology is that of a 3-torus. This means that there are
$(N-1)(M-1)(L-1)$ independent vertices. We also require $L,M$ and $N$
to be even numbers.
%%%%%%%%%%%%%%%%%%%%%%%%%%%%%%%%%%%%%%%%%%%%%%%%%%%%%%%%%%%%%%%%%%%%%%
It is easy to see that
for any particular constraint ${\chi}_{I}$, there exists another
constraint ${\chi}_{J}$ such that
$ \{ {\chi}_{I},{\chi}_{J} \} $ does not weakly vanish.
Therefore, all constraints ${\chi}_{I}$ are second class constraints.
We now follow the procedure first set forth by Dirac in dealing with
second class constraints.  We define the Dirac bracket of two
functions
on the phase space, $\xi$ and $\psi$, as
\f
\{ \xi,\psi \}_{db} = \{ \xi,\psi \}_{pb} - \{ \xi,{\chi}_{I} \}_{pb}
{\Omega}^{IJ} \{ {\chi}_{J},\psi \}_{pb}
\ff
where ${\Omega}^{IJ}$ is defined by
\f
\{ {\chi}_{I},{\chi}_{J} \}_{pb} {\Omega}^{JK} = {\delta}^{K}_{I}
\ff
Therefore, the Dirac bracket of any function on the phase space
with any ${\chi}_{I}$ vanishes by construction.

It is not difficult to see that the resulting Dirac brackets are
non-local, in the sense that an $a^i$ on one face of the lattice
can have a nonvanishing Dirac bracket with an $e^i$ on a link
arbitrarily far from it.  This happens because to find the Dirac
brackets we have to invert the matrix $\{\chi_I , \chi_J \}$.  This
may seem unphysical, but it is actually necessary if the Dirac algebra
of the $a^i$'s and their conjugate momenta (defined by 33) is to
be local.  The problem is that the relation between the $e^i$'s
and the conjugate momenta are nonlocal. (Actually, a natural rearrangement
of the configuration variables makes the Dirac bracket local; see below.)

The straightforward way to proceed is to write a computer program to
invert the relation (37) and find the Dirac bracket for any given lattice,
which we did.  However, a small trick gives us an easy form of $\Omega$.  We
rearrange the constraints
\f
0 = {\chi}_{a^i_{\vec{n}\hat{c}}} = {\Pi}_{a^i_{\vec{n}\hat{c}}} -
{d{\cal L} \over d{\dot{a}}^i_{\vec{n}\hat{c}}}
\ff
in such a way that all new constraints contain only one $e^i_{
\vec{n}\hat{a}}$, and the rest in the same manner.

Using the above rearrangement of constraints, (37) is simple enough
to solve analytically.  The Dirac bracket is
\begin{eqnarray}
&&  {\{ a^j_{\vec{m} \hat{d}} ,e^i_{\vec{n} \hat{a}} \}}_{db}=\label{a,e}\\
&&(-1)^{k+l+m-1}(sign(k)-\delta_{k,0})(sign(l)+\delta_{l,0})\times
\nonumber\\
&&\times(sign(m)+\delta_{m,0})\delta_{\hat{a},\hat{d}} \delta_{i,j}\nonumber\\
&&
 \vec{m}-\vec{n} = k\hat{a}+l\hat{b}+m\hat{c};\;\;\;\;\;\;\;\;\;\;
\hat{a}\cdot \hat{b}=\hat{a}\cdot \hat{c}=\hat{b}\cdot \hat{c}=0\nonumber
\end{eqnarray}
where
\f
sign(l)=0\;\;for\;\; l=0,\;\; +1\;\; for\;\; l>0,\;\; -1\;\;
  for\;\; l<0 \label{sign}
\ff
This
expression was checked by directly performing the inversion required to
solve (37) on a computer using a 6x6x6 cubic lattice.

This expression seems nonlocal at first glance.  However, if we rearrange
the configuration variables as
\begin{eqnarray}
{e'}^i_{\vec{n} \hat{a}}&=&e^i_{\vec{n}-\hat{a}, \hat{a}}+e^i_{\vec{n} \hat{a}}
\label{e'}\\
{a'}^i_{\vec{n} \hat{a}}&=&a^i_{\vec{n} \hat{a}} +\sum_{\hat{b},\hat{c}\neq
\hat{a}\;\;\hat{b}\neq \hat{c}}
\frac{1}{2}a^i_{({\vec{n}-( \hat{b}+\hat{c})) \hat{a}} }+
a^i_{({\vec{n}-\hat{b}) \hat{a}} }\label{a'}
\end{eqnarray}
(this will not change the number of configuration variables), we get in
fact a local Dirac bracket, see also fig(\ref{local}):
\f
{\{ {a'}^j_{\vec{n} \hat{b}},{e'}^i_{\vec{m} \hat{a}}
\}}_{db}=\delta_{\hat{a},\hat{b}} \delta_{i,j}
\delta_{n,m}\label{l}
\ff
%%%%%%%%%%%%%%%%%%%% ADDED %%%%%%%%%%%%%%%%%%%%%%%%%%%%%%%%%%%%%%%%%%%
\begin{figure}[t]
\begin{picture}(600,70)(30,0)
\Huge
\put(105,35){$\{ \;\;\;\;\;\;\;\;\;\;\; , \;\;\;\;\;\;\;\;\; \}
=\delta_{\vec{n},\vec{m}}$}
\put(140,40){\circle*{5}}
\put(140,20){$\vec{n}$}
\put(215,40){\vector(1,0){25}}
\put(190,40){\vector(1,0){25}}
\put(215,40){\circle*{5}}
\put(215,20){$\vec{m}$}
\put(125,5){\line(0,1){50}}
\put(140,15){\line(0,1){50}}
\put(155,25){\line(0,1){50}}
\put(125,5){\line(3,2){30}}
\put(125,30){\line(3,2){30}}
\put(125,55){\line(3,2){30}}
\normalsize
\end{picture}
\begin{center}\parbox{12cm}{
\caption{The Dirac brackets are local for a certain combination of
our original discrete configuration variables.}
\label{local}}\end{center}\end{figure}
%%%%%%%%%%%%%%%%%%%%%%%%%%%%%%%%%%%%%%%%%%%%%%%%%%%%%%%%%%%%%%%%%%%%%%

We now turn to the problem of finding the algebra of the constraints.
To simplify the job of finding the
Dirac bracket algebra of the constraints, we considered first an
approximation in which we
neglect all non-abelian terms in the constraints, so that
\f
{\cal C}_{\vec{n}\hat{a}}= \epsilon_{\hat{a}\hat{b}\hat{c}}
 e^i_{\vec{n} \hat{b}} b^i_{\vec{n} \hat{c}}
\ff
\f
{\cal C}_{\vec{n}} = \epsilon_{ijk} \epsilon_{\hat{a}\hat{b}\hat{c}}
e^i_{\vec{n} \hat{a}}  e^j_{\vec{n} \hat{b}} b^k_{\vec{n} \hat{c}}
\ff
\f
{\cal G}^i_{\vec{n}} = \sum_{\hat{a}}
\left (
e^i_{\vec{n}\hat{a}} - e^i_{(\vec{n}-\hat{a})\hat{a}}
\right )
\ff
are the vector, scalar, and Gauss constraints with
\f
b^i_{\vec{n}\hat{a}} = \sum_{\hat{b},\hat{c} \neq \hat{a}}
\epsilon^{\hat{a}\hat{b}\hat{c}}
\left ( a^i_{\vec{n}\hat{c}} - a^i_{(\vec{n}-\hat{b})\hat{c}}
\right )
\ff

As mentioned in the introduction, this approximation corresponds,
in the continuum theory, to the limit in which Newton's constant
is taken to zero, as described in \cite{Gtozero}.
%%%%%%%%%%%%%%%%%%%%%%%% ADDED %%%%%%%%%%%%%%%%%%%%%%%%%%%%%%%%%
Note that both Bianchi identities (\ref{Bianchi}) and (\ref{Bianchi2}) are
 now fulfilled, as there
are no cubic terms at all.

The results of the Dirac bracket algebra is now as follows:
The algebra of a generic lattice (as specified above) {\em does not close}.
Almost all dirac brackets between the vector and scalar constraints
can't be expressed as a linear combination of constraints.
This is obvious when computing a particular
bracket and has been checked in a linearization program in the case of
a $6\times6\times6$ lattice. Treated as (secondary) constraints,
the non vanishing dirac brackets actually are first class.
The algebra must eventually close, as there are
a finite number of combinations of $a^i_{\vec{n}\hat{a}}$'s and
$e^i_{\vec{n}\hat{a}}$'s.
The full algebra including the brackets of the
new first class constraints is quite complicated for a generic lattice.
Therefore, we present the full algebra for a
simple but not trivial lattice and only a part for the generic lattice.

Let us start with the simple case. For a $2\times 2\times N$ lattice,
meaning $N-1$ independent vertices
on a line, much is simplified. Let the line be along the $x$-axes. Then,
there are only $N-1$ constraints left (as $b_{\vec{n}x}=0$ which means
that $e_{\vec{n}x}=0$), namely
\f
C_{\vec{n}x}\equiv
D_{\vec{n}}=e^i_{\vec{n}y}b^i_{\vec{n}z}-e^i_{\vec{n}z}b^i_{\vec{n}y}
\;\;\;for\;n\;=1,2...N
\ff
Now,
\begin{eqnarray}
\{ D_{\vec{n}}, D_{\vec{n}+m\hat{x}} \} & = &
(-1)^{m-1} sign(m)(
e^i_{\vec{n}y}b^i_{\vec{m}z}-e^i_{\vec{n}z}b^i_{\vec{m}y}+
e^i_{\vec{m}y}b^i_{\vec{n}z}-e^i_{\vec{m}z}b^i_{\vec{n}y})\\
&&\equiv  {\cal{D}}_{(\vec{n},\vec{n}+m\hat{x}) }\nonumber
\end{eqnarray}
, where $sign(l)$ is defined in (\ref{sign}). The full algebra then becomes
\begin{eqnarray}
\{ D_{(\vec{n}+l\hat{x},\vec{n}+m\hat{x})},
 D_{(\vec{n}+p\hat{x},\vec{n}+q\hat{x})} \} =
(-1)^{l-p-1}sign(l-p)D_{(\vec{n}+l\hat{x},\vec{n}+p\hat{x})}+\\
+(-1)^{l-q-1}sign(l-q)D_{(\vec{n}+l\hat{x},\vec{n}+q\hat{x})}+\\
+(-1)^{m-p-1}sign(m-p)D_{(\vec{n}+m\hat{x},\vec{n}+p\hat{x})}+\\
+(-1)^{m-q-1}sign(m-q)D_{(\vec{n}+m\hat{x},\vec{n}+q\hat{x})}
\end{eqnarray}
where
\f
2D_{\vec{n}} \equiv D_{(\vec{n},\vec{n})}
\ff
Among the $({N \over 2})$ number of $D_{(\vec{n},\vec{m})}$'s there are only
one relation,
\f
\sum_{\vec{n},\vec{m}} D_{(\vec{n},\vec{m})}=0
\ff

In case of a generic $(even)\times (even)\times (even)$ lattice with opposite
sides
identified, we give
the following sample of the full algebra
\f
\{ {\cal{C}}_{\vec{n} \hat{a}},  {\cal{C}}_{(\vec{n}+l\hat{a})\hat{a}} \}
=(-1)^{l-1} sign(l) \varepsilon_{\hat{a}\hat{b}\hat{c}}
 e^i_{(\vec{n} \hat{b}} b^i_{\vec{m}) \hat{c}} \equiv
(-1)^{l-1} sign(l) {\cal{C}}_{(\vec{n},\vec{n}
+l\hat{a})
\hat{a}},
\label{vip}
\ff
where parentheses denotes symmetrization.

It is possible to interpret the non-vanishing dirac brackets (\ref{vip}) if
we treat them as constraints. They are then, as we have seen, first class
and we may write the ${\cal{C}}_{(\vec{n},\vec{m})\hat{a}}$'s using (\ref{fi})
as
\f
{\cal{C}}_{(\vec{n},\vec{m})\hat{a}} =
(\phi_{ij}(x_{\vec{n}})-\phi_{ji}(x_{\vec{m}}))
\varepsilon_{\hat{a}\hat{b}\hat{c}}b^i_{\vec{n}\hat{b}}b^j_{\vec{m}\hat{c}}
\label{interp}
\ff
we realize that $\phi_{ij}(x_{\vec{n}})$ must be the same for all $\vec{n}$'s
if the
${\cal{C}}_{(\vec{n},\vec{m})\hat{a}}$'s are
to vanish,
\f
\phi_{ij}(x_{\vec{n}})=\phi_{ij}(x_{\vec{m}})\;\;for\;all\;\vec{n}\;and\;\vec{m}
\ff
Hence, all vertices are identical, there is only {\em{one}} independent vertex.
That is, {\em if} we consider the ''non-closedness of the original algebra'' as
 constraints.

%%%%%%%%%%%%%%%%%%%%%%%%%%%%%%%%%%%%%%%%%%%%%%%%%%%%%%%%%%%%%%%%

\section{A four dimensional formulation}

In the past two sections we worked out some details of a Hamiltonian
formulation of a discretization of general relativity based on
distributional
fields of the form of (2).  In this section we would like to sketch out a
fully
four dimensional lagrangian formulation which differs from the one
we
have just developed in that time, as well as space, will be treated
discretely.

To do this we introduce into a four dimensional manifold $\cal M$,
representing
spacetime, an  arbitrary hypercubic lattice.  By extending the
definitions (2) and (15) we will require that the self-dual curvature
has support on
the faces of the lattice, and the self-dual connection has support on
the
three dimensional surfaces of the lattice.  We will see that the result
of
these assumptions is a natural reduction
of the Lagrangian theory derived from the Capovilla-Dell-Jacobson
form
(6) to a Lagrangian form of a theory with a countable set of degrees
of
freedom.

We will follow what we did in the three dimensional case and
label the sites of the four dimensional lattice by $x_{\vec{n}}$,
the links by $\gamma_{\vec{n} \hat{a}}$, the faces by
${\cal S}_{\vec{n}\hat{a}\hat{b}}$ and the three dimensional
surfaces
by ${\cal R}_{\vec{n}\hat{a}\hat{b}\hat{c}}$.  All of the indices are
now
four dimensional and refer to
the object that is gotten by moving in the indicated positive
directions
from the site $\vec{n}$.  Also, the labels of the two and
three dimensional surfaces are assumed to be antisymmetric in
the indices.  Assuming a hypercubic structure, it is easy
to label the faces making up the boundaries of the three dimensional
surfaces, they are
\f
\partial {\cal R}_{\vec{n}\hat{a}\hat{b}\hat{c}} =
{\cal S}_{\vec{n}\hat{a}\hat{b}} \cup
{\cal S}_{\vec{n}+\hat{c},\hat{a}\hat{b}} \cup
{\cal S}_{\vec{n}\hat{a}\hat{c}} \cup
{\cal S}_{\vec{n}+\hat{b},\hat{a}\hat{c}} \cup
{\cal S}_{\vec{n}\hat{b}\hat{c}} \cup
{\cal S}_{\vec{n}+\hat{a},\hat{b}\hat{c}}
\ff

With these labellings, we then introduce a Lie algebra valued one
form
that has support on the three dimensional surfaces of the lattice.  It
is
parametrized by a Lie algebra element
$a^i_{\vec{n}\hat{a}\hat{b}\hat{c}}$
attached to each three dimensional surface,
${\cal R}_{\vec{n}\hat{a}\hat{b}\hat{c}}$, and it is written as,
\f
A_a^i(x) \equiv {1 \over G} \sum_{\vec{n}\hat{a}\hat{b}\hat{c}} \int
d^3{\cal R}_{\vec{n}\hat{a}\hat{b}\hat{c}}^{bcd}(\rho )
\epsilon_{abcd}
\delta^4 \left (
x, {\cal R}_{\vec{n}\hat{a}\hat{b}\hat{c}}(\rho )  \right )
a^i_{\vec{n}\hat{a}\hat{b}\hat{c}}.
\ff
Here, $a,b,c,d$ are four dimensional spacetime indices and $\rho$ are
three
coordinates on the surface.

Following the derivation of (18) it is straightforward to show that the
curvature
associated with this connection is well defined and distributional, and
has
support on the two dimensional faces of the lattice.  It is written as,
\f
F_{ab}^i (x) = {1 \over G} \sum_{\vec{n}\hat{a}\hat{b}}
\int d^2 {\cal S}_{\vec{n}\hat{a}\hat{b}}^{cd}(\sigma )
\epsilon_{abcd}
\delta^4 \left ( x, {\cal S}_{\vec{n}\hat{a}\hat{b}}   (\sigma )
\right )
b^i_{\vec{n}\hat{a}\hat{b}}  ,\label{47}
\ff
where the Lie algebra elements $b^i_{\vec{n}\hat{a}\hat{b}} $
associated
to each face of the lattice are given by
\f
b^i_{\vec{n}\hat{a}\hat{b}} = \sum_{\hat{c} \neq \hat{a},\hat{b}}
\left (    a^i_{\vec{n}\hat{a}\hat{b}\hat{c}} -
a^i_{\vec{n}-\hat{c},\hat{a}\hat{b}\hat{c}}          \right ) +
{G \over 16} \epsilon^{ijk}\sum_{\hat{c}\hat{d}}
\epsilon_{\hat{a}\hat{b}\hat{c}\hat{d}}
\left [  a^j_{\vec{n}\hat{a}\hat{b}\hat{c}}   +
a^j_{\vec{n}-\hat{c}, \hat{a}\hat{b}\hat{c}}  \right ]
\left [  a^k_{\vec{n}\hat{a}\hat{b}\hat{d}}   +
a^k_{\vec{n}-\hat{d}, \hat{a}\hat{b}\hat{d}}  \right ]  .
\ff

It is now straightforward to plug (\ref{47}) into the CDJ form of the action
(6) and find that the reduced action is
\f
S[ a,\phi ] =
\sum_{\vec{n}\hat{a}\hat{b}\hat{c}\hat{d}}
\epsilon_{\hat{a}\hat{b}\hat{c}\hat{d}} \
\bar{b}^i_{\vec{n}\hat{a}\hat{b}} \ \bar{b}^j_{\vec{n}\hat{c}\hat{d}}
\
\phi^{-1}_{\vec{n} \ ij} .\label{49}
\ff
Here $\phi_{\vec{n}}^{ij}= \phi (x_{\vec{n}})^{ij}$ is the matrix of
scalar
fields that satisfies (7) and (8) and $\bar{b}^i_{\vec{n}\hat{a}\hat{b}}$ is
the average of the $b^i_{\vec{n}\hat{a}\hat{b}}$'s on the four
surfaces
in the $\hat{a}\hat{b}$ plane that touch the site $\vec{n}$,
\f
\bar{b}^i_{\vec{n}\hat{a}\hat{b}} \equiv {1\over 4}
\left [ b^i_{\vec{n}\hat{a}\hat{b}} + b^i_{\vec{n}-\hat{a},
\hat{a}\hat{b}}
+ b^i_{\vec{n}-\hat{b}, \hat{a}\hat{b}}
+ b^i_{\vec{n}-\hat{a}-\hat{b}, \hat{a}\hat{b}}
\right ]
\ff

\section{Conclusions}

In this  paper we have introduced two new
approximation schemes for general relativity which are
based on reductions to spaces of solutions based on
finite dimensional spaces of solutions.  We believe that
it is likely that either the Lagrangian or the Hamiltonian
formulation proposed here could
be used to provide arbitrary good approximations
to solutions to Einstein's equations.  However, there are
a number of things that need to be checked in order to insure
that either leads to a useful approximation scheme for either
classical or quantum gravity.   We would like to close by
listing what we think remains to
be done to establish the usefulness
of the formulations introduced here.

1)  The study of the algebra of the full constraints must be completed.
It is not necessary that the algebra be first class for a reduction such
as the one given here to provide a useful approximation method; in
the case that the algebra is second class the reduction must be
understood
as constituting a partial gauge fixing.  However, to go ahead, it is
important
to know what the algebra is.

2)  The reality conditions must be formulated for the reduced theory,
in both the Hamiltonian and Lagrangian case.

3)  It may be more convenient, for some purposes, to use a
simplicial rather than a cubic or hypercubic lattice.  In this
case it may be useful to introduce a dual lattice, whose faces
are to be in one to one correspondence with the lines of the
lattice (in the three dimensional case.)  The faces of this dual
lattice may then be assigned areas which, according to the
results described in section 4 are given by $|e_I|$, where
$I$ labels the faces (and the lines of the original lattice.)
The simplices of this dual lattice can then be taken to constitute
a piecewise flat Regge manifold whose edge lengths are determined
from the areas of the faces.  In this way one can reformulate
the system described here as a canonical
formulation of the Regge calculus in which the three metric variables
are treated conventionally, but the conjugate
variables are the distributional
self-dual curvature and connections, which
live on a lattice dual to the
simplicial lattice.

Furthermore, the Lagrangian formulation for simplicial lattices then
appears to be a generalization of the four dimensional topological
quantum field theories of Ooguri\cite{ooguri} and
Crane and Yetter\cite{craneyetter}.    This is a connection that
deserves further exploration.

4)  It is interesting to speculate the extent to which this formulation
could be useful for some approach to quantum gravity.  It does
seems unlikely that a direct quantization of the hamiltonian system
described in section 2 could be useful, given that direct imposition
of the diffeomorphism constraints in the continuum already reduces
quantum gravity to something very much like a finite system.
On the other hand, it may be that the Lagrangian formulation described
in section 4 could be the  basis for a path integral quantization.
Especially in the light of the recent impressive
progress in four dimensional Monte Carlo simulations
involving simplicial approximations, it would
be interesting to investigate the duality suggested by the
previous remarks between
the present formulation and a Regge calculus formulation.  It may
also be that a direct attack on the path integral, given the action
(\ref{49}) may yield interesting results.  As in any path integral formulation,
the key problem that must  be solved before reliable results can
be extracted from the path integral is to find the correct measure
of integration.

It is clear that a great deal of work remains to be done to establish
whether or not the formulations proposed here may be useful
for practical calculations in classical or quantum gravity.
We hope that the results established here suffice to justify
this further work.

\section*{Appendix.  Observables and the meaning of distributional
geometries}

In this appendix we would like to return to
the question of how distributional geometries
of the kind we have been discussing can be used to approximate
arbitrarily
well smooth geometries.   The idea of using such distributional
geometries to approximate smooth geometries arose during work on
the loop representation approach
to quantum gravity\cite{weaves}, when it was realized
that the classical limits of loop states could be associated with
distributional geometries of the form (2).  It was then discovered
that certain classical observables, which are functions of the
classical fields, could be extended unambigously so that they were
defined on an extension of the phase space of general relativity
that includes distributional fields of the form of (2) and (16).
Now, this extension cannot be made for all observables, indeed,
local observables, such as the metric at a point, cannot be given
an unambiguous meaning for the distributional geometries.  However,
there are other observables, which are non-local which can be
defined in such a way that they are meaningful when evaluated on
the distributional geometries.

In references \cite{weaves,review-lee,review-Abhay} one can
find discussions of the quantum operators corresponding to
these quantities.  Very similar considerations apply to the
extension of the observables to classical distributional geometries,
as these results have never appeared we describe here the
details for one observable, which is the area observable.

The main idea is to use a distributional geometry to
approximate a smooth
geometry by mimicing the weave construction of the quantum
theory.
For example, if one wants to approximate a metric, $q_{ab}$, which
is slowly varying on a scale $L$, up to an accuracy of $a << L$, one
can do the following.  We require a set of curves $\alpha$, such that
the distributional geometry $\tilde{E}_\alpha^a$, defined by (1),
satisfies the following requirement:

For every surface $\cal S$, whose area, given the metric $q_{ab}$,
which we will call ${\cal A}[{\cal S},q]$ is larger than $L^2$,
and whose extrinsic curvature is bounded by $1/L^2$,
we require that
\f
\left | {\cal A}[{\cal S},E_\alpha ]-{\cal A}[{\cal S},q]   \right |
< {a^2 \over {\cal A}[{\cal S},q]}
\ff
If this is the case then we say that the distributional geometry
$\tilde{E}^a_\alpha$ area-approximates the metric $q_{ab}$
on the scale $a$.

We may note that the problem of approximating
a smooth geometry by a distributional one is closely connected to
the problem of taking the classical limit of a loop state of the
quantum theory and showing that it
approximates a classical metric\cite{review-carlo,review-lee}.
These two problems are
closely related because the loop states are eigenstates of
certain operators that measure the three metric, such that the
eigenvalues involve the distributional frame field associated
to a loop by (1).  Thus, the problem of extending an observable
from smooth to distributional geometries is closely related to
the problem of constructing good operators for those same
quantities in the quantum theory through a regularization
procedure.

We would thus like, in the remainder of this section, to show how
the area  operator can be extended to the distributional
geometries.  From the results it will be clear that it is easy
to solve the problem of constructing distributional geometries that
area-approximate any smooth geometry arbitrary well.  Analogous
results hold for other observables, including those involving the
self-dual connections and frame fields, but we do not give them
explicitly here.

Let us consider, then, an arbitrary two dimensional surface in
$M$, which we will denote
$\cal S$.  The coordinates of $\cal S$  are given by $S^a
(\sigma_\alpha )$,
where the two $\sigma_\alpha$, $\alpha =1,2$ are coordinates
on the surface.  Let us the consider the  observable ${\cal A}[{\cal
S} ]$,
which assigns to each surface $\cal S$ its area induced from the
three metric on
$M$.  One would not normally think that this observable could
be extended to distributional geometries.   However, we will now
show that because our distributional frame fields are also
densities,
${\cal A}[{\cal S}]$ can   be defined by
a certain procedure, which allows it to extend unambiguously to
distributional loop geometries.  The result of this will be that,
despite the fact
that distributional geometries of the form of (2),
are defined only on a set of measure zero, the observables
${\cal A}[{\cal S}]$ are, when nonzero, finite.  They assign
finite areas to surfaces that cross the loop $\alpha$.

We begin by writing the usual expression for ${\cal A}[{\cal S}]$
in the case of a smooth, nondegenerate geometry,
\f
 {\cal A}[{\cal S}]  = \int_{\cal S} \sqrt{h} ,\label{52}
\ff
where $h$ is the determinant of the induced two metric,
$h_{ab}=q_{ab} -n_a n_b $, where $n^a $ is the unit normal.  A
simple
calculation show that $h= \tilde{\tilde{q}}^{ab}n_a n_b$.  Now,
it is not hard to show  that $\tilde{\tilde{q}}^{ab}$ cannot itself
be extended to distributional loop geometries.  This can be
demonstrated by a direct extension of the argument for the non-existence
of a unambigous renormalized operator for $\tilde{\tilde{q}}^{ab}$
in the loop representation\cite{weaves,review-carlo,review-lee}.
As a result, we must construct
the area (\ref{52}) through a limit that does not need this extension.  To
do this, let us divide the surface up into   $N$ disjoint regions
${\cal S}_i$,
such that ${\cal S} =\cup_i {\cal S}_i$.  We then have
 \f
 {\cal A}[{\cal S}] = \sum_i {\cal A}[{\cal S }_i]
\ff
 We will proceed by introducing an approximation for the square of
${\cal A}[{\cal S}_i] $ which becomes exact in the limit of
infinitesimal surfaces.  This is,
\f
{\cal A}_{approx}^2[{\cal S}_i] \equiv \int d^2S^{ab}_i
\int d^2 S^{\prime \ cd}_i T^{**}(S,S^{\prime} )_{ab \ cd }
\ff
where
$T^{**}(x,y )_{ab \ cd } \equiv \epsilon_{abe} \epsilon_{cdf}
T^{ef}(x,y)$.
To show that this approximates the area of the surface element
for small surfaces, we use the facts that in the limit
$T^{ab}(S,S^{\prime} )  \approx \tilde{\tilde{q}}^{ab}$.  We may
invert the relation $h=\tilde{\tilde{q}}^{ab}n_a n_b$ to find that
$\tilde{\tilde{q}}^{ab}=hn^an^b-r^{ab}$ where $r^{ab}n_b=0$.
An infinitesimal element of
area is given by $d{\cal A}= d^2S^{ab}\sqrt{h}n^a \epsilon_{abc}$,
from which it follows that,
\f
d{\cal A}^2 = d^2S^{ab} d^2S^{cd} \epsilon_{abe}\epsilon_{cdf}
\tilde{\tilde{q}}^{ef}\label{54}
\ff
For smooth fields, this is then equal to (\ref{52}) in the limit of small
areas.
We may then consider the limit in which we divide the surface up
into smaller and smaller elements, so that $N \rightarrow \infty$.
It then follows that,
\f
{\cal A}[ {\cal S}] = \lim_{N \rightarrow \infty} \sum_{i=1}^N
\sqrt{{\cal A}_{approx}^2[{\cal S}_i]} .\label{56}
\ff
For smooth, nondegenerate metrics, this is a long way round to go
to define the area.  But, as we will now show,   this
particular definition extends
to classical distributional loop geometries.

We then evaluate (\ref{54}) for a distributional loop geometry given by
(1).
We find,
\begin{eqnarray}
{\cal A}_{approx}^2[{\cal S}_i] &=& a^4 \int d^2 S_i^{ab} \oint
d\alpha^c (s)
\delta^3( S, \alpha (s)) \epsilon_{abc}
\int d^2 S_i^{\prime \ de} \oint d\alpha^f (t)
\delta^3( S^{\prime} , \alpha (t)) \epsilon_{def}
\nonumber \\
&&\times
Tr\left [ w (s)U_{\gamma_{S,S^{\prime}}} (0,\pi) w(t)
U_{\gamma_{S,S^{\prime}}} (\pi,2 \pi)  \right ]
\end{eqnarray}
The expression
\f
\int d^2 S_i^{ab} \oint d\alpha^c
\delta^3( S_i \alpha ) \epsilon_{abc}
=I[{\cal S}_i, \alpha ]
\ff
is the intersection number of the curve with the surface.  It is zero
unless they intersect, in which case it is equal to $\pm 1$
depending
on the orientations.  Now, as we take the limit of $N \rightarrow
\infty$
we will pass a point at which the absolute value of the
intersection
number of $\alpha$ with each surface element is at most one.  At
that
point we have
\f
{\cal A}_{approx}^2[{\cal S}_i] = \left (
a^2 |w(s^*_i)|I[{\cal S}, \alpha ]    \right )^2
\ff
where $s^*_i$ is the intersection point of the curve with the
$i$'th surface element.
At this point further subdivisions do not change the value of the
sum in (\ref{56}), so the limit is equal to
\f
{\cal A}[{\cal S}] = a^2 \sum_i |w(s^*_i )|
\ff
where the sum is over the intersection points of the curve with the
whole surface.

Given this result, it is clear how to construct a curve $\alpha$
such that the distributional geometry $\tilde{E}_\alpha^a$
area-approximates a given smooth metric $q_{ab}$.  Indeed,
the construction
mimicks the one that has been already given in the quantum
case \cite{weaves,review-carlo,review-lee};
the main idea is to arrange the loops so that the
sum of the contributions in (\ref{56}) is equal to the area of each surface.

\section*{ACKNOWLEDGEMENTS}

L.S. and M.M. would like to thank Warner Miller and members of
the Syracuse
relativity group for useful suggestions and comments about this
work.  It was supported by the National Science Foundation
under grants PHY90-16733 and INT88-15209 and by research funds
provided by Syracuse University.
%%%%%%%%%%%%%%%%%%%%%% ADDED %%%%%%%%%%%%%%%%%%%%%%%%%%%%%%%%%%%%%%%%%%%
O.B. wishes to thank Ingemar Bengtsson, Gabriele Ferretti, Joakim Hallin,
Stephen Hwang
and G{\"o}ran Ericsson for useful discussions and comments.
%%%%%%%%%%%%%%%%%%%%%%%%%%%%%%%%%%%%%%%%%%%%%%%%%%%%%%%%%%%%%%

\end{document}